\documentclass[aip,prb,twocolumn,amsmath,amssymb, reprint]{revtex4-1}

\usepackage{graphicx}
\usepackage{dcolumn}
\usepackage{bm}
\usepackage{textcomp}
\makeatletter

\newcommand{\Rmnum}[1]{\expandafter\@slowromancap\romannumeral #1@}
\makeatother

\date{\today}

\begin{document}

\title[] {Using light and heat to controllably switch and reset disorder configuration in nanoscale devices}

\author{A. M. See}
\affiliation{School of Physics, University of New South Wales,
Sydney NSW 2052, Australia}
\author{M. Aagesen}
\author{P. E. Lindelof}
\affiliation{Nanoscience center, University of Copenhagen,
Universitetsparken 5, DK-2100 Copenhagen, Denmark}
\author{A. R. Hamilton}
\author{A. P. Micolich}
\email{adam.micolich@nanoelectronics.physics.unsw.edu.au}
\affiliation{School of Physics, University of New South Wales,
Sydney NSW 2052, Australia}

\begin{abstract}
Quantum dots exhibit reproducible conductance fluctuations at low
temperatures due to electron quantum interference. The sensitivity
of these fluctuations to the underlying disorder potential has only
recently been fully realized. We exploit this sensitivity to obtain
a novel tool for better understanding the role that background
impurities play in the electrical properties of high-mobility
AlGaAs/GaAs heterostructures and nanoscale devices. In particular,
we report the remarkable ability to first alter the disorder
potential in an undoped AlGaAs/GaAs heterostructure by optical
illumination and then reset it back to its initial configuration by
room temperature thermal cycling in the dark. We attribute this
behavior to a mixture of C background impurities acting as shallow
acceptors and deep trapping by Si impurities. This `alter and reset'
capability, not possible in modulation-doped heterostructures,
offers an exciting route to studying how scattering from even small
densities of charged impurities influences the properties of
nanoscale semiconductor devices.
\end{abstract}

\maketitle

Disorder is an important issue in nanoelectronics; as a device is
reduced in size it becomes more sensitive to temporal fluctuations
and spatial inhomogeneities in the charged impurity
distribution.\cite{SimmonsNP08} The temporal fluctuations cause
decoherence\cite{HuPRL06,PeterssonPRL10}, noise\cite{KurdakPRB97}
and device irreproducibility.\cite{YangAPL09, ScannellPRB12} These
are troublesome for the development of quantum applications for
semiconductor devices. Spatial inhomogeneities interrupt electron
flow\cite{JuraNP07} adversely affecting\cite{PanPRL11} practical
realization of concepts such as topological quantum computation
using the $5/2$ fractional quantum Hall state\cite{NayakRMP08}.
These barriers to applications have fueled research to reduce,
control and better understand disorder in nanoscale electronic
devices.

Approaches to disorder reduction for nanoscale devices include
modulation-doping\cite{DingleAPL78}, short period superlattice
doping\cite{FriedlandPRL96, UmanskyJCG09}, and undoped
heterostructures where the carriers are induced using either a
metal\cite{HirayamaJAP96, HarrellAPL99} or degenerately-doped
semiconductor gate\cite{SolomonEDL84, KaneAPL93}. While
modulation-doped structures still provide the highest mobilities,
undoped heterostructures provide two distinct advantages. First,
short range neutral disorder dominates over long range Coulombic
disorder. This brings benefits such as enhanced robustness of the
$5/2$ fractional quantum Hall state,\cite{PanPRL11} and improved
experimental access to the metallic state generated by
electron-electron interactions in 2D systems.\cite{ClarkeNP08}
Second, the absence of intentional ionized impurities gives
electrical properties that are remarkably robust to thermal
cycling\cite{SeePRL12}, in stark contrast to modulation-doped
heterostructures\cite{ScannellPRB12}. Both features demonstrate
there is much more to disorder than the popular metric of mobility
alone.\cite{MicolichFdP13} They also strongly motivate the quest for
a deeper understanding of the nature of disorder in undoped
heterostructure devices.

We recently developed a novel approach to studying disorder in
nanoscale devices. It relies on the fact that a quantum dot's low
temperature magnetoconductance shows quantum interference
fluctuations that are highly sensitive to the paths electrons take
in traversing the device.\cite{FengPRL86} These `magnetoconductance
fluctuations' (MCF) are influenced by both the dot
geometry\cite{MarcusPRL92} and the underlying disorder
potential\cite{JiPRB95, MicolichFdP13} in dots smaller than the
large-angle scattering length, i.e., in the `ballistic' transport
regime. If the dot geometry is fixed, changes in the MCF can be used
to probe the physics of dopants, e.g., the intentional Si
donors\cite{ScannellPRB12} and acceptors,\cite{CarradPRB14} within
the device. We recently made a remarkable finding -- the MCF becomes
highly reproducible, even after thermal cycling to room temperature,
if intentional dopants are removed, i.e., the quantum dot is made
using an undoped heterostructure.\cite{SeePRL12} This result
naturally leads to interesting questions: An undoped heterostructure
is never $100\%$ pure, so what role does the low density of
non-intentional `background' impurities inevitably incorporated into
the heterostructure during growth play? Since these impurities are
`invisible' thermally, can we probe them another way, for example,
optically? Might it be possible to use this small density of
impurities as a new way to control electrical properties towards
novel applications?

Here we demonstrate the ability to alter the disorder potential in
undoped heterostructures by optical illumination, detect the
resulting changes in spatial charge configuration by looking for
changes in low temperature magnetoconductance, and then reset the
disorder potential back to its initial configuration by thermally
cycling the device to room temperature. This remarkable capability,
not possible in conventional modulation-doped
heterostructures,\cite{ScannellPRB12} provides a significant
opportunity to better understand how the spatial configuration of
even small densities of charged impurities influences transport in
nanoscale devices. Note that the technique could be readily extended
towards materials beyond III-V semiconductors, e.g.,
graphene,\cite{StaleyPRB08,UjiieJPCM09} one only needs a quantum dot
showing quantum interference fluctuations.

\begin{figure}
\includegraphics[width=7cm]{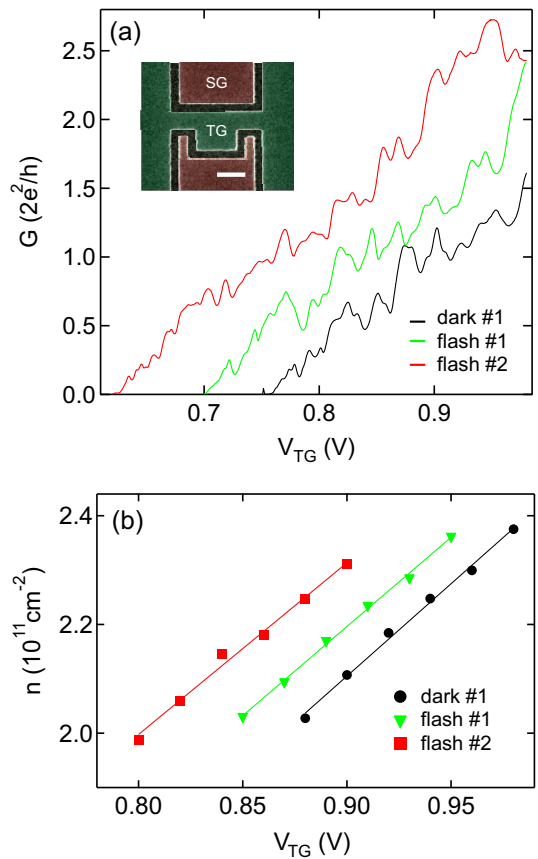}
\caption{The device and how the gate characteristics and density
change with illumination. (a) A comparison of dot conductance $G$ vs
top-gate voltage $V_{TG}$ obtained at temperature $T = 250$~mK and
side-gate voltage $V_{SG} = 0$~V. The three traces shown are:
(black) prior to illumination denoted `dark $\#1$', (green) after
first illumination denoted `flash $\#1$', and (red) after second
illumination denoted `flash $\#2$'. (Inset) Scanning electron
micrograph of the device; the scale bar represents $500$~nm. The
top-gate (TG) is shaded green, the side-gates (SG) are shaded red.
(b) Two-dimensional electron density $n$ vs. $V_{TG}$ for different
amounts of illumination. The straight lines are fits to the
experimental data.}
\end{figure}

A scanning electron micrograph of our device is shown inset to
Fig.~1(a). It was made using a heterostructure consisting of an
undoped GaAs substrate overgrown with $160$~nm undoped AlGaAs,
$25$~nm undoped GaAs, and a $35$~nm n$^{+}$ GaAs cap, which remains
highly conductive at low temperature. The cap is divided into three
independent gates -- a top-gate (green) and two side-gates (red) --
using electron-beam lithography and shallow wet etching. The NiGeAu
ohmic contacts are produced using an established self-aligned
process.\cite{KaneAPL93, SeeAPL10} The dot and the 2D electron gas
(2DEG) source and drain reservoirs are populated electrostatically
when the top-gate (TG) is biased to a sufficiently positive voltage
$V_{TG}$. The width of the quantum point contacts (QPCs) connecting
the dot to source and drain can be tuned via a negative voltage
$V_{SG}$ applied to the side-gates (SG). Electrical characterization
of a Hall bar made using the same heterostructure at 4.2~K gave a
mobility $\mu \sim 300,000$~cm$^{2}$/Vs at $n \sim 1.8 \times
10^{11}$~cm$^{-2}$, corresponding to $\ell \sim 2.1~\mu$m, which is
larger than the dot dimensions ($760 \times 660$~nm). Electrical
measurements were performed at temperature $T \approx 250$~mK using
a pumped $^3$He cryostat (Oxford Instruments Heliox), with the
magnetoconductance $G(B)$ obtained by standard four-terminal lock-in
techniques using a fixed $100~\mu$V excitation voltage at a
frequency of $11$~Hz. The variable magnetic field $B$ was applied
along the heterostructure growth direction (perpendicular to the
plane of Fig.~1(a) inset). Illumination was performed using a red
LED (see Fig.~S1 for emission spectrum) mounted in close proximity
to the device. Illumination is implemented via a control box that
drives an LED current of $1$~mA for a single $100$~ms pulse. Further
details on the undoped devices and electrical measurements are
available elsewhere.\cite{KaneAPL93, SeeAPL10, SeePRL12}

Figure~1(a) shows traces of zero field conductance $G(B = 0)$ versus
$V_{TG}$ obtained in a single cooldown: before illumination denoted
`dark $\#1$' (red), after a $100$~ms illumination denoted `flash
$\#1$' (black), and after another $100$~ms illumination denoted
`flash $\#2$' (green). In each case $G$ increases as $V_{TG}$ is
made more positive, reflecting a gate-induced accumulation of
electrons in the dot and source/drain contacts. The threshold
voltage $V_{th}$ -- defined here as the lowest $V_{TG}$ where $G$
becomes non-zero -- decreases from $0.76$~V to $0.7$~V and then
$0.62$~V after the first and second illuminations, respectively.
This $V_{th}$ shift indicates the optical ionization of background
impurities and a commensurate increase in electron density. To more
clearly demonstrate the density increase in Fig.~1(b) we plot the
electron density $n$ from Shubnikov-de Haas measurements of an
adjacent Hall-bar segment on this device versus $V_{TG}$; the red
circles, black triangles and green squares are data obtained before,
between and after the two illuminations, respectively. With each
illumination, the density increases by $4-5\%$. This density
increase is persistent, as previously shown in an inverted, undoped
AlGaAs/GaAs heterointerface.\cite{SakuJJAP98} The slope of $n$
versus $V_{TG}$ in Fig.~1(b) is constant, demonstrating that
illumination does not alter the gate-2DEG capacitance.

The fluctuations in the gate sweeps in Fig.~1(a) arises from quantum
interference\cite{BirdPRL99} and represents a `fingerprint' of
transport through the dot similar to $G(B)$. The exact fluctuation
pattern varies with each illumination due to a change in the
disorder potential within the dot. We can assess this further using
$G(B)$,\cite{ScannellPRB12,SeePRL12} but doing this rigorously
requires caution. If we simply compare $G(B)$ at a given $V_{TG}$
before, between and after illuminations, the disorder potential
change is masked by an associated density-driven change in Fermi
wavelength $\lambda_{F} \sim n^{-\frac{1}{2}}$ (for completeness, we
nonetheless present this comparison in Fig.~S2); such a comparison
is only meaningful at common $n$. To properly demonstrate the
illumination-induced change in disorder potential we use the
following procedure. First, we plot the gate sweeps to a transformed
gate voltage axis $V^{\prime}_{TG} = V_{TG} - V_{th}$ as shown in
Fig.~2(a). The three traces have a common gradient but different
fluctuation patterns such that the traces cross each other at
various points. At these crossing points the device has the same
$G(B = 0)$ and $n$ for a given $V^{\prime}_{TG}$ before, between and
after the two illuminations, providing the most fair basis for
comparing the corresponding $G(B)$ traces. One such instance is
highlighted by the arrow at $V^{\prime}_{TG} = +0.90$~V in Fig.~2(a)
inset; the three corresponding $G(B)$ traces are plotted in
Fig.~2(b). The $G(B)$ traces vary significantly confirming a change
in the dot's disorder potential.\cite{ScannellPRB12, SeePRL12} This
is highlighted in Fig.~2(c), where we plot the difference $\delta G
= G(B)_{flash~\#1} - G(B)_{dark~\#1}$ (solid), $G(B)_{flash~\#2} -
G(B)_{flash~\#1}$ (dashed), or $G(B)_{flash~\#2} - G(B)_{dark~\#1}$
(dotted) versus $B$. The latter highlights the cumulative change
arising from the two illuminations. The cumulative rms fluctuation
across the $0 < B < 0.5$~T range is $0.076$, $0.088$ and $0.108
\times 2e^{2}/h$, respectively. Each illumination causes an
approximately equivalent change, but this effect ultimately
diminishes due to the finite population of light-active centers;
this typically occurs after four $100$~ms illumination pulses for
this device.

\begin{figure}
\includegraphics[width=7cm]{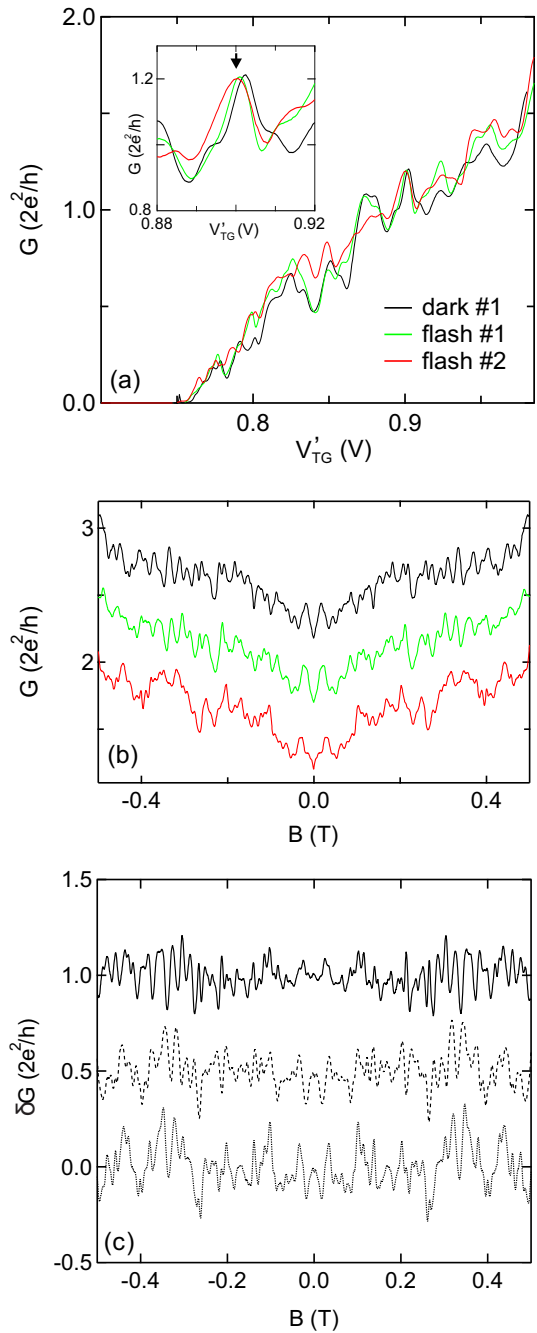}
\caption{Demonstration that illumination changes the dot's disorder
potential. (a) Comparison of $G$ vs transformed top gate voltage
$V^{\prime}_{TG} = V_{TG} - V_{th}$ where $V_{th} = 0, 55, 130$~mV
for the dark $\#1$ (black), flash $\#1$ (green) and flash $\#2$
(red) traces, respectively, using data from Fig.~1(a). (Inset) A
magnified view of the main panel centered at $V^{\prime}_{TG} =
0.90$~V showing a crossing point for the three traces as indicated
by the arrow. (b) Dark $\#1$, flash $\#1$ and flash $\#2$ $G$ vs $B$
traces obtained at $V_{TG} = 0.900$~V (black), $0.845$~V (green) and
$0.770$~V (red) corresponding to $V^{\prime}_{TG} = 0.90$~V. (c) The
MCF difference $\delta G = G(B)_{flash~\#1} - G(B)_{dark~\#1}$
(solid), $G(B)_{flash~\#2} - G(B)_{flash~\#1}$ (dashed), or
$G(B)_{flash~\#2} - G(B)_{dark~\#1}$ (dotted) vs $B$ highlighting
the $G(B)$ changes resulting from illumination. Traces in (b) and
(c) are offset sequentially downwards by $0.5\times 2e^{2}/h$ for
clarity.}
\end{figure}

The most remarkable aspect of this device emerges in response to
thermal cycling to room temperature. Figure~3(a) shows $G(B = 0)$
versus $V_{TG}$ from the second cooldown before illumination,
denoted dark $\#2$ (solid), along with the corresponding trace from
the first cooldown dark $\#1$ (dashed). These two traces are
essentially identical, as are $G(B)$ traces obtained at $V_{TG} =
+0.9$~V (Fig. 3(b)). Note that $V_{TG} = +0.9$~V is chosen out of
convenience here; matching $G(B)$ for various $V_{TG}$ as reported
previously.~\cite{SeePRL12} The $G(B)$ similarity is clear in the
difference trace $\delta G(B) = G(B)_{dark~\#2} - G(B)_{dark~\#1}$
in Fig.~3(c), which is presented at matching scale to Fig.~2(c) for
direct comparison. The cumulative rms fluctuation for the trace in
Fig.~3(c) is $0.030 \times 2e^{2}/h$, a factor of $2-4$ smaller than
for the traces in Fig.~2(c). This indicates that the disorder
potential returns to its initial configuration upon thermal cycling.
In other words, in our undoped devices, illumination can be used to
randomly reconfigure the disorder, and it can be `reset' to a
repeatable base configuration by thermal cycling.

An interesting question is: What is the nature of the disorder
enabling this ability to optically alter and then thermally reset
the disorder potential? Two features of the data provide important
clues. The first is that illumination increases the electron density
(Fig.~1(b)), i.e., the sample displays persistent positive
photoconductivity.\cite{LangPRL77, NelsonAPL77, StormerAPL81} The
second is that the impurity population returns to its initial
ionization configuration after thermal cycling (Fig.~3(a/b)). There
are two possible impurity models that satisfy these features: Si
background impurities acting as deep traps and C background
impurities acting as shallow acceptors; both may exist at different
densities. We first address these individually with respect to
persistent photoconductivity, before considering how they fit with
our observation that we can reset the MCF, and thereby the disorder
potential, to its initial configuration by room temperature thermal
cycling.

\begin{figure}
\includegraphics[width=6.5cm]{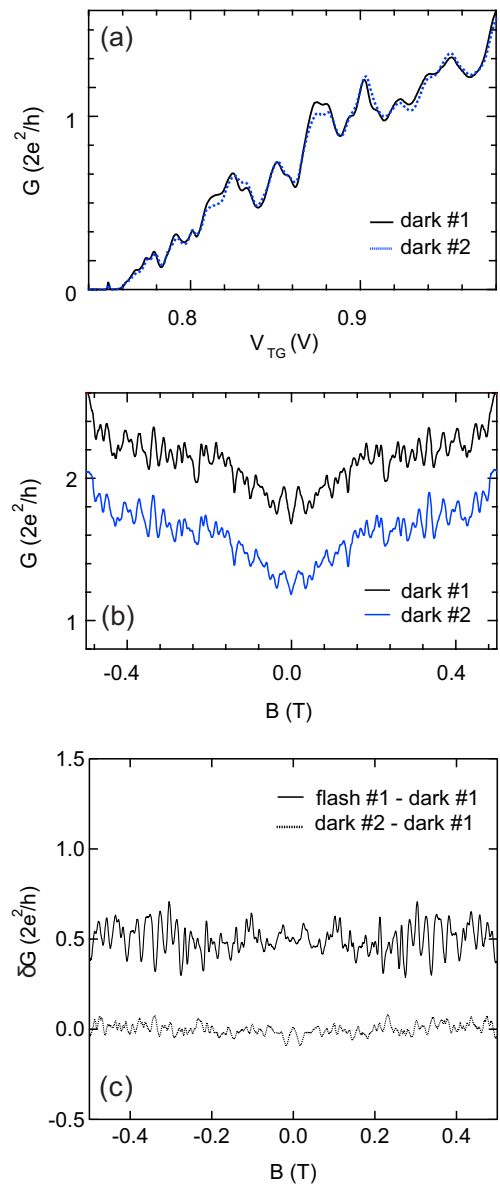}
\caption{Demonstration of the ability to reset the disorder
potential to its initial configuration with thermal cycling. (a)
Comparison of $G$ vs $V_{TG}$ obtained before illumination on the
first cooldown, denoted dark $\#1$ (black solid line), and after two
illuminations and room temperature thermal cycling in the dark,
denoted dark $\#2$ (blue dashed line). (b) Comparison of $G$ vs. $B$
at $V_{TG}$ = $0.9$~V for dark $\#1$ (black) and dark $\#2$ (blue);
the dark $\#2$ trace is offset vertically by $0.5\times 2e^{2}/h$
for clarity. The red trace is offset upwards by $0.5\times 2e^{2}/h$
for clarity. (c) The MCF difference $\delta G = G(B)_{dark~\#2} -
G(B)_{dark~\#1}$ vs $B$ (dashed) along with $\delta G =
G(B)_{flash~\#1} - G(B)_{dark~\#1}$ vs $B$ (solid) from Fig.~2(c)
for comparison. The solid trace is offset upwards by $0.5\times
2e^{2}/h$ for clarity.}
\end{figure}

A well known cause of persistent positive photoconductivity in
Al$_{x}$Ga$_{1-x}$As/GaAs heterostructures with $x~>~0.2$ are
deep-trapping Si DX centers in the AlGaAs layer.\cite{LangPRB79,
MooneyJAP90} Briefly, the Si DX center can take three possible
states: shallow hydrogenic donor states d$^{0}$ and d$^{+}$, and a
deep trap DX$^{-}$ that is stabilized by lattice
deformation.\cite{MooneyJAP90, BuksSST94} The ground-state is
DX$^{-}$, it can be optically excited to release one or both of its
trapped electrons to become d$^{0}$ or d$^{+}$.\cite{MooneySST91,
BuksSST94} The latter is the ultimate outcome either way, as the
transition d$^{0}~\rightarrow~$d$^{+}$ requires much less energy
than DX$^{-}~\rightarrow~$d$^{0}$.\cite{MooneySST91} The optically
liberated electrons either join the 2DEG or are swept into the
n$^{+}$ cap by the gate electric field. Note that these liberated
electrons cannot be retrapped by Si impurities unless the sample is
warmed above $100$~K due to a lattice-deformation-induced energy
barrier surrounding these impurities.\cite{LangPRB79} Thus an
associated change in the charge between gate and 2DEG occurs,
producing a negative shift in $V_{th}$, and equivalently, an
increase in $n$ at fixed $V_{TG}$, i.e., persistent
photoconductivity. The presence of Si background impurities in our
device are a logical possibility because a) the 2DEG is immediately
adjacent to a $160$~nm undoped Al$_{0.33}$Ga$_{0.67}$As layer and b)
the MBE chamber contains a Si source for degenerate doping of the
GaAs cap in these structures and modulation-doping of other
heterostructures grown during the same chamber evacuation. Ge, S,
Se, Sn and Te impurities can also produce deep-trapping DX centers
in AlGaAs;\cite{LangPRB79, KumagaiAPL84} in particular, S is a
common impurity in As MBE sources\cite{SkrommeJAP85, LarkinsJCG87}
and is proposed to also show DX behavior in GaAs.\cite{ParkPRB96,
DuPRB05} Note that Si is a shallow hydrogenic donor in GaAs with
states d$^{0}$ and d$^{+}$ only.

C background impurities are also common in MBE-grown AlGaAs/GaAs
heterostructures.\cite{SkrommeJAP85, LarkinsJCG87} Carbon is not
known to show DX behavior,\cite{LangPRB79} acting instead as a
shallow acceptor.\cite{GianniniJAP93} However, an ionized C acceptor
a$^{-}$ may be neutralized to a$^{0}$ by hole-capture from a nearby
photo-generated electron-hole pair. This electron-hole pair
generation should occur, particularly in our GaAs layers, since we
use a red LED in this experiment.\cite{FletcherPRB90} This process
should result in positive persistent photoconductivity providing the
photo-generated electron enters the 2DEG or n$^{+}$-GaAs cap, rather
than being captured to re-ionize the same acceptor. This is possible
because, unlike the DX center, there is no
lattice-deformation-induced barrier surrounding the C impurity.
Illumination can decrease the two-dimensional hole gas density in
C-doped AlGaAs/GaAs heterostructures,\cite{GerlJCG07} presumably via
a similar mechanism. One advantage of this model is that it is
viable for both the undoped AlGaAs spacer and the GaAs substrate,
unlike the Si DX center model discussed above.

We now turn our attention to the ability to reset the disorder
potential and MCF to the initial, pre-illumination configuration by
room temperature thermal cycling. The fact that we observe this
means that upon cool-down all impurities return to their initial
charge states. In the simplest case, this implies: a) all acceptors,
e.g., C impurities, are in the ionized a$^{-}$ state, b) all donors,
e.g., Si, S, etc. in GaAs are in the d$^{+}$ state, and c) all
donors, e.g., Si, S, etc. in the AlGaAs are in the DX$^{-}$ state.
While this ignores some exotic possibilities, e.g., impurity
complexes, these are likely rare due to the very low impurity
density. We now consider the plausibility of this full ionization
scenario. As a shallow acceptor, C in GaAs commonly shows electrical
activation efficiencies well above $80\%$\cite{GianniniJAP93}, with
$100\%$ activation found for doping densities less than $3 \times
10^{17}$~cm$^{-3}$.\cite{ItoAPL93} Activation is enhanced in
Al$_{x}$Ga$_{1-x}$As, increasing with Al content
$x$.\cite{GianniniJAP93} We also expect full ionization of Si
shallow donors in GaAs as their activation energy is $6$~meV,
compared to $26$~meV for C shallow acceptors.\cite{PierretBook02}
Complete Si activation is indeed observed at low doping
densities;\cite{SchubertAPL94} autocompensation due to Si dopant
amphotericity\cite{SchupplerAPL93} and more complex donor
deactivation mechanisms\cite{DomkePRB96} only become important for
Si densities above $10^{19}$~cm$^{-3}$. The return of all Si
impurities in the AlGaAs to the DX$^{-}$ state is the most
unexpected outcome because in modulation-doped devices this does not
occur.\cite{BuksSST94, ScannellPRB12} However, the intentional Si
density in modulation-doped structures is $1,000-2,000\times$
greater than the background impurity density in these undoped
heterostructures,\cite{MacLeodPRB09} even if we assume that all of
the background impurities are Si, which is highly unlikely. In this
instance, the free electron density in the AlGaAs during cool-down
may sufficiently exceed the Si impurity density to doubly-occupy all
DX centers -- there may be two coinciding mechanisms that aid this
process. The first is the tendency for background impurities,
particularly Si, to ride upwards on the AlGaAs growth
front.\cite{HeiblumJVSTB85, PfeifferAPL91} This would push the bulk
of the Si background impurities closer to the n$^{+}$-GaAs cap. The
second is that the Fermi energy is pinned against the conduction
band edge in the n$^{+}$-GaAs cap, which raises the local Fermi
level in the upper parts of the AlGaAs spacer (e.g. see Fig.~2 of
Ref.~\cite{ClarkeJAP06}). This should enhance the free electron
density in the AlGaAs layer, increasing the probability for Si
background impurities to take the DX$^{-}$ state. Additionally, the
lower Si background impurity density would increase the impurity
spacing by a factor of at least $10$ relative to modulation-doped
heterostructures. This large spacing $\gtrsim 65$~nm should
dramatically reduce charge migration between adjacent DX centers
(i.e., DX$^{-} + $d$^{+} \rightarrow$ d$^{+} +$ DX$^{-}$), further
enhancing MCF stability. It should also prevent charge migration
between acceptor sites, and the associated slow drift in $G(B)$
recently reported for hole quantum dots.\cite{CarradPRB14}

\begin{figure}
\includegraphics[width=7cm]{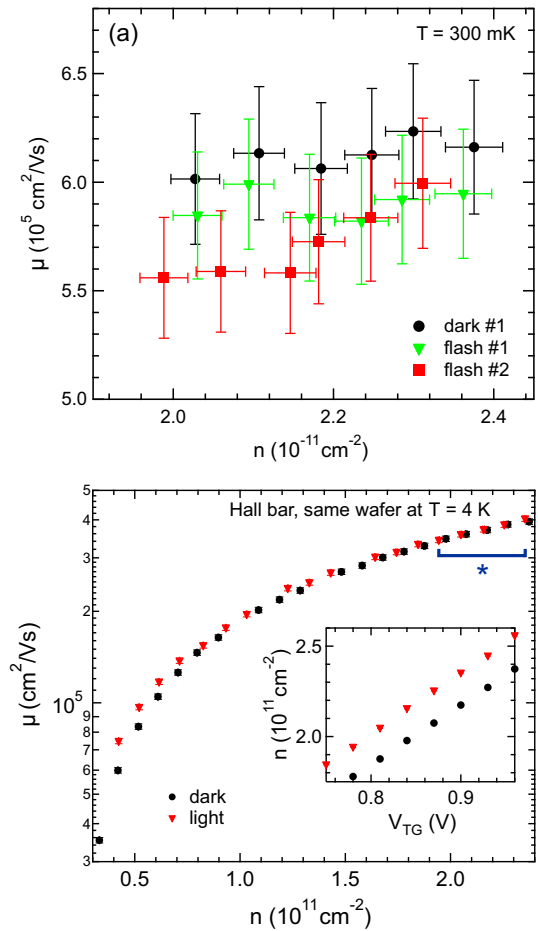}
\caption{How illumination influences the mobility (a) The mobility
$\mu$ vs electron density $n$ for dark $\#1$ (black circles), flash
$\#1$ (green triangles) and flash $\#2$ (red squares) measured from
Shubnikov-de Haas oscillations measured on a Hall-bar segment
immediately adjacent to the quantum dot used to obtain data in
Figs.~1-3. Data was obtained at $T = 300$~mK from a region of the
Hall-bar adjacent to the quantum dot. (b) The mobility $\mu$ vs
electron density $n$ obtained before (black circles) and after
illumination (red triangles) measured for a separate Hall-bar sample
made using the same heterostructure. Only the error bars in $\mu ~
2\%$ are shown, the error bars in $n ~ 0.2\%$. The data was obtained
at $T = 4$~K and the blue bracket marked $*$ indicates the operating
density range for our quantum dot sample (cf. Fig.~1(b)). Inset
shows $n$ vs top-gate voltage $V_{TG}$ for this Hall bar for
comparison with data from the quantum dot sample in Fig.~1(b).}
\end{figure}

Turning to the mobility, the transition a$^{-} \rightarrow$ a$^{0}$
should result in a mobility increase as it eliminates a charged
scattering center. In contrast, there should be little change in the
mobility from the transition DX$^{-} \rightarrow$ d$^{+}$ because
negative DX$^{-}$ centers and positive d$^{+}$ donors both cause
Coulomb scattering of electrons in the 2DEG. This argument ignores
DX center correlation effects,\cite{BuksSST94} which might be
justified given the $\sim10\times$ higher donor separation; but for
completeness, were there any correlations the random nature of DX
center photoionization should reduce them, causing $\mu$ to decrease
slightly.\cite{BuksSST94} Figure~4(a) shows $\mu$ vs $n$ obtained
Shubnikov-de Haas measurements from a Hall bar segment immediately
adjacent to the quantum dot used for Figs.~1-3. This data suggests
that $\mu(n)$ decreases with illumination, but the trend is barely
significant relative to measurement error. In Fig.~4(b) we present
characterization data for a separate Hall bar from the same
heterostructure without any etched quantum dot. Here we observe an
increase in $\mu(n)$ with illumination, which ranges from over
$20\%$ at low $n$ to almost zero at high $n > 2 \times
10^{15}$~cm$^{-2}$ where our device is operated (indicated by the
blue $*$ in Fig.~4(b)). A similar increase in $\mu(n)$ was
previously reported for undoped AlGaAs/GaAs heterostructures by Saku
{\it et al.}\cite{SakuJJAP98}, but their increase was much larger,
decreasing from $\sim 46\%$ at $n \sim 2 \times 10^{10}$~cm$^{-2}$
to $\sim 25\%$ at $n \sim 2 \times 10^{11}$~cm$^{-2}$. Given the
lack of a clear shift in $\mu$ with illumination in our operating
$n$ range, we suspect that both mechanisms a$^{-} \rightarrow$
a$^{0}$ and DX$^{-} \rightarrow$ d$^{+}$ are at play in our device,
perhaps with the a$^{-} \rightarrow$ a$^{0}$ process being slightly
dominant. Additional studies using custom doped heterostructures
and/or allied analytical methods, e.g. deep level transient
spectroscopy (DLTS),\cite{LangJAP74} may assist further in
understanding the physics of the underlying dopant photoexcitation
processes involved in the device behavior we report.

In summary, we have shown the ability to alter the disorder
potential in undoped heterostructures by optical illumination, and
then reset the disorder potential back to its initial configuration
by thermally cycling the device to room temperature. Our data
suggests that this process likely arises from a mixture of two
processes: a) photoexcitation of ionized C background acceptors back
to their neutral state, and b) photoionization of Si DX centers.
This remarkable `alter and reset' capability is not possible in
conventional modulation-doped heterostructures\cite{ScannellPRB12}
and offers an exciting new route to studying how spatial aspects of
scattering from even small densities of charged impurities
influences transport in nanoscale semiconductor devices. Our
approach also has great potential as a tool for optimising
semiconductor materials for applications where extremely low
disorder/scattering is vital to operation, e.g., quantum information
processing,\cite{UmanskyJCG09, PanPRL11} including materials beyond
III-V semiconductors, e.g., graphene.\cite{StaleyPRB08, UjiieJPCM09}


This work was funded by the Australian Research Council (ARC). APM
acknowledges an ARC Future Fellowship and ARH acknowledges an ARC
Outstanding Researcher Award. We thank L.~Eaves, D.J.~Carrad and
S.P.~Bremner for helpful discussions. This work was performed in
part using the NSW node of the Australian National Fabrication
Facility (ANFF).

\end{document}